\documentclass[twocolumn,prd,floatfix,nofootinbib,superscriptaddress,longbibliography,notitlepage]{revtex4-1}

\pdfoutput=1

\usepackage{amsmath, amssymb, amsfonts, amsthm, latexsym, mathrsfs, bbm, slashed}

\usepackage[usenames,dvipsnames]{xcolor}
\usepackage{graphicx}

\graphicspath{{../calculations/}{./figs/}}

\usepackage{xifthen}

\usepackage[inline]{enumitem}

\usepackage{setspace}
\usepackage[marginal, multiple]{footmisc}

\usepackage[T1]{fontenc}
\usepackage[utf8]{inputenc}
\usepackage{lmodern}

\usepackage[unicode, colorlinks, allcolors=blue!70!black, linktocpage, pdfusetitle]{hyperref}
\definecolor{CiteColor}{rgb}{0,0.5,0}
\hypersetup{citecolor=CiteColor}
\definecolor{RefColor}{rgb}{0.55,0,0}
\hypersetup{linkcolor=RefColor}
\usepackage[all]{hypcap}

\usepackage{orcidlink}

\usepackage{verbatim}

\usepackage{cancel}

\usepackage{microtype}

\usepackage{standalone}
\usepackage{tikz}
\usetikzlibrary{arrows.meta}
\usetikzlibrary{decorations.markings}
\ifdefined\myext
\usetikzlibrary{external}
\fi

\newcommand{\nn}{\nonumber}

\newcommand{\pd}{\partial}
\newcommand{\cd}{\nabla}

\DeclareMathOperator{\sign}{sign}
\DeclareMathOperator{\diag}{diag}

\newcommand{\E}{\mathcal{E}}
\renewcommand{\L}{\mathcal{L}_z}
\newcommand{\Q}{\mathcal{Q}}

\renewcommand{\min}{\text{min}}
\renewcommand{\max}{\text{max}}
\newcommand{\isco}{\text{isco}}
\newcommand{\ibco}{\text{ibco}}
\newcommand{\isso}{\text{isso}}
\newcommand{\ibso}{\text{ibso}}

\newcommand{\equat}{\text{equat}}

\newcommand{\psep}[1][]{p_{\text{sep}}\ifthenelse{\isempty{#1}}{}^{#1}}
\newcommand{\esep}[1][]{e_{\text{sep}}\ifthenelse{\isempty{#1}}{}^{#1}}
\newcommand{\xsep}[1][]{x_{\text{sep}}\ifthenelse{\isempty{#1}}{}^{#1}}

\newcommand{\reg}{\text{reg}}

\newcommand{\figoverview}{%
\begin{figure*}[t]
  \begin{center}
    \includegraphics[width=\textwidth]{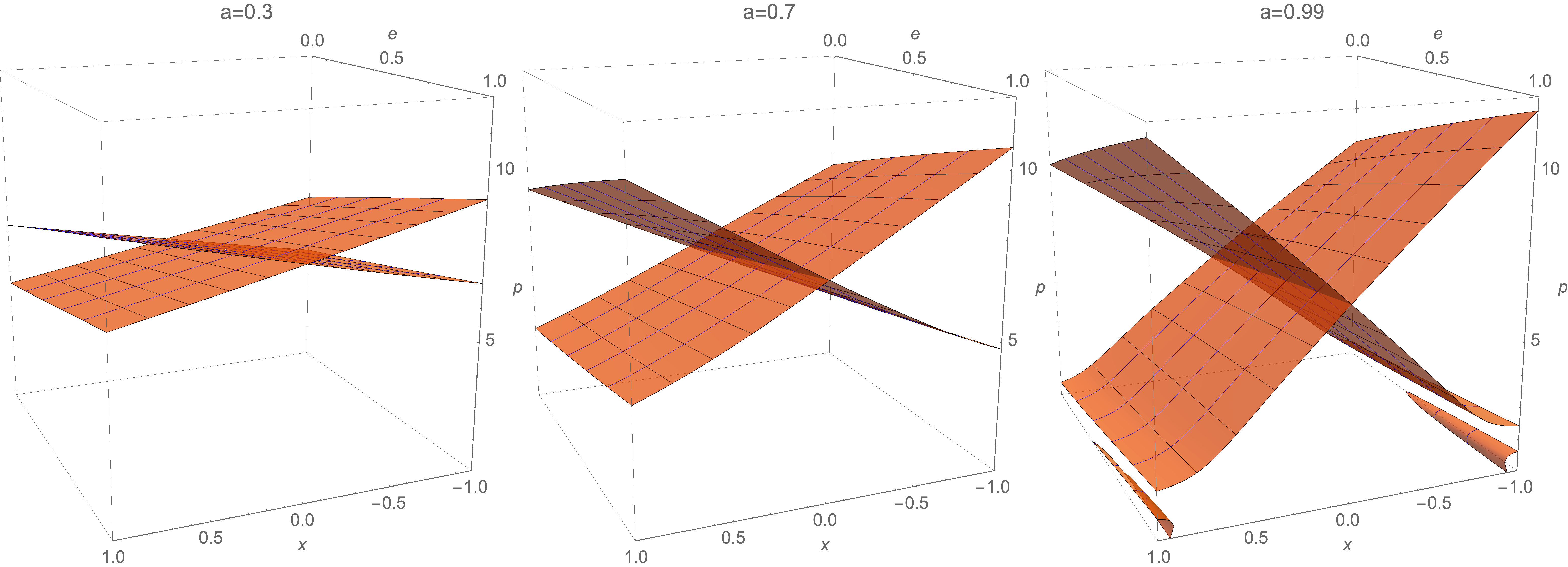}
  \end{center}
  \caption{%
    Overview of solutions of the separatrix polynomial $S=0$ at
    selected values of spin.  At each value of $(a,e,x)$, there are
    solutions for both the prograde and retrograde values of the
    separatrix $\psep$ for the corresponding value of $x^{2}$.  There
    may also be unphysical solutions, as seen in the rightmost panel,
    appearing at smaller values of $p$.  The physical branch is the
    one sloping ``downward'' in $x$, i.e.~$\psep$ decreases as $x$
    increases.
  }
  \label{fig:overview}
\end{figure*}
}

\begin{document}

\title{The location of the last stable orbit in Kerr spacetime}
	
\author{Leo C.~Stein\,\orcidlink{0000-0001-7559-9597}}
\email{lcstein@olemiss.edu}
\affiliation{Department of Physics and Astronomy,
  The University of Mississippi, University, MS 38677, USA}
	
\author{Niels Warburton\,\orcidlink{0000-0003-0914-8645}}
\email{niels.warburton@ucd.ie}
\affiliation{School of Mathematics and Statistics,
  University College Dublin, Belfield, Dublin 4, Ireland.}

\hypersetup{pdfauthor={Stein and Warburton}}

\begin{abstract}
  Black hole spacetimes, like the Kerr spacetime, admit both stable
  and plunging orbits, separated in parameter space by the separatrix.
  Determining the location of the separatrix is of fundamental
  interest in understanding black holes, and is of crucial importance
  for modeling extreme mass-ratio inspirals.  Previous numerical
  approaches to locating the Kerr separatrix were not always efficient
  or stable across all of parameter space.  In this paper we show that
  the Kerr separatrix is the zero set of a single polynomial in
  parameter space.  This gives two main results.  First, we thoroughly
  analyze special cases (extreme Kerr, polar orbits, etc.), finding
  strict bounds on the limits of roots, and unifying a number of
  results in the literature.  Second, we pose a stable numerical
  method which is guaranteed to quickly and robustly converge to the
  separatrix.  This new approach is implemented in the Black Hole
  Perturbation Toolkit, and results in a $\sim 45\times$
  speedup over the prior robust approach.
\end{abstract}
	
\maketitle

\section{Introduction}
\label{sec:intro}

The existence of unstable and plunging orbits for test body motion in
general relativity is one of the key differences in celestial
mechanics between Newtonian and Einstein gravity.  In the strong
gravitational field around black holes, a region of the parameter
space appears where stable bound orbits are no longer possible.  In
this region test bodies either plunge directly into the black hole or
are on unstable orbits to which any slight perturbation will trigger
the body to plunge.  This has important consequences in astrophysics.
For example, the inner edge of a black hole accretion disk is set by
the location of the innermost stable circular orbit (ISCO).  The
relation between the ISCO radius and the black hole's spin is
exploited to make measurements of the rotation rate of astrophysical
black holes~\cite{McClintock:2011zq}.

The ISCO delineates one edge of a more general structure called the
`separatrix' that divides the stable region of the parameter space
from the unstable/plunging region.  This separatrix is particularly
important for the physics of extreme mass-ratio inspirals
(EMRIs)~\cite{Babak:2017tow, Berry:2019wgg}, key sources for the future space-based
gravitational wave detector LISA.  The event rate of these binaries is
strongly influenced by the location of the separatrix, with highly
spinning massive black holes more likely to capture stellar mass
compact objects on prograde orbits~\cite{AmaroSeoane:2012cr}.  Once
the secondary is captured its orbit will decay through gravitational
wave emission until it reaches the separatrix and plunges into the
massive black hole.  Consequently, knowledge of the location of the
separatrix is a key ingredient in models of these
binaries~\cite{Babak:2006uv, Chua:2017ujo, Ori:2000zn,
  OShaughnessy:2002tbu, Sundararajan:2008bw, Apte:2019txp,
  Burke:2019yek, Compere:2019cqe}.  The region of parameter space near
the separatrix is also interesting as it is here that the well known
relativistic orbital precession is taken to the extreme, with
arbitrary large precession possible when approaching the
separatrix~\cite{Glampedakis:2002ya}.

Calculating the location of the separatrix for generic orbits that
could be eccentric or inclined is non-trivial.  For certain limiting
cases the location can be found analytically, but in general numerical
solutions must be found.  There are a variety of methods in the
literature~\cite{Sundararajan:2008bw, AmaroSeoane:2012cr} to find the
separatrix for generic orbits but these are not always efficient or
stable across the entire parameter space.
In this work we show that the separatrix is an algebraic variety,
and
derive a \emph{single} polynomial, of degrees
$(12,12,12,4)$ in the indeterminates $(a,p,e,x^{2})$, which are the
orbital parameters detailed below.
The roots of this polynomial give the location of the separatrix.
This has two benefits:
(i) it is easy to analyze the limiting cases (equatorial motion, extreme Kerr, etc.), and
(ii) we can apply rapidly convergent methods for finding the roots of polynomials.
We analyze many limits and detail several numerical schemes, with a full
implementation provided in the Black Hole Perturbation
Toolkit~\cite{BHPToolkit}.

The organization of this paper is as follows. 
Sec.~\ref{sec:geod-orbit-param} discusses time-like geodesic motion in Kerr spacetime focussing on bound orbits. 
Sec.~\ref{sec:separatrix} defines the separatrix and other special orbits and derives the separatrix polynomial. 
We look at solutions to the separatrix polynomial in interesting limiting cases in Sec.~\ref{sec:limits}. 
Finally, in Sec.~\ref{sec:num-impl} we discuss numerical methods for solving the separatrix polynomial. 
In the appendices we give some additional details, including results for special orbits such as the innermost bound spherical orbit.
We also present an alternative robust method for numerically locating the separatrix in Appendix~\ref{apdx:homoclinic_method}.
Throughout this article we use geometrized units such that the speed of light and the gravitational constant are equal to unity.
We also use standard Boyer-Lindquist coordinates $(t,r,\theta,\varphi)$ and use the metric signature $(-~+~+~+)$.

\section{Time-like geodesics in\\* Kerr spacetime}
\label{sec:geod-orbit-param}

Given any spacetime, let us denote the trajectory of a timelike
(non-spinning) test body of mass $\mu$ by a curve $x^\alpha(\tau)$
where $\tau$ is the proper time as measured along the world line.
The four-velocity of the body is given by $u^\alpha = dx^\alpha/d\tau$ where for timelike motion we have (with our choice of metric signature) $u^\alpha u_\alpha = -1$.
The test body's trajectory is governed by the second-order differential equation $u^\beta \nabla _\beta u^\alpha = 0$ where $\nabla_\beta$ is the covariant derivative with respect to the background geometry.

Hereafter we focus on motion about a Kerr black hole.
The Kerr spacetime is parameterized by the black hole mass, $M$, and its spin $a$, where $a = J/M$ with $J$ the angular momentum of the black hole.
We choose $J\ge 0$ so that $0\le a \le M$.
For motion about a Kerr black hole the Killing symmetries of the spacetime give rise to conserved quantities. 
Two of these, the orbital energy and (azimuthal) angular momentum, are
associated with isometries of the metric, with associated Killing
vector fields $(\pd_{t})^{\alpha}$ and $(\pd_{\varphi})^{\alpha}$.
The third, the Carter constant, is related to a hidden symmetry
associated with a Killing tensor $\Q^{\alpha\beta}$ of the spacetime,
satisfying $\cd_{(\alpha}\Q_{\beta\gamma)}=0$.
With these constants of motion, and the conserved mass of the test body, the geodesic equations in Boyer-Lindquist coordinates can be written in first-order form:
\begin{align}
  \Sigma^2 \left(\frac{dr}{d\tau}\right)^2 &= R(r)\\
  \Sigma^2 \left(\frac{d\theta}{d\tau}\right)^2&= \Theta(\theta)\\
  \Sigma \frac{d\varphi}{d\tau}					&= \frac{a}{\Delta}(2r\E - a\L) + \frac{\L}{\sin^2\theta}		\\
  \Sigma \frac{dt}{d\tau}							&= \frac{(r^2 + a^2)^2\E - 2 a r \L}{\Delta} - a^2 \E \sin^2\theta
  \,,
\end{align}
where $\Sigma \equiv r^2 + a^2\cos^2\theta$, $\Delta \equiv r^2 - 2Mr + a^2 $ and
\begin{align}
  R(r) ={}& -\beta r^4 + 2 r^3 - (a^2\beta+\L^2) r^2 \nonumber \\
  &+ 2(a\E - \L)^2 r - \Q\Delta\label{eq:radial_potential}\\
  \Theta(\theta) ={}& \Q - \cos^2\theta\left\{a^2\beta +\frac{\L^2}{\sin^2\theta}\right\}
  \,,
\end{align}
where $\beta = (1-\E^2)$. In the above equations and hereafter $\E,
\L$, and $\Q$ denote the specific energy, angular momentum and
Carter constant, respectively.  These are related to the tangent
$u_{\alpha}=\mu^{-1} p_{\alpha}$ and the Killing vectors and tensor via
\begin{align}
  \E &\equiv - (\pd_{t})^{\alpha} u_{\alpha} \,, &
  \L &\equiv  (\pd_{\varphi})^{\alpha} u_{\alpha} \,, &
  \Q &\equiv  \Q^{\alpha\beta} u_{\alpha} u_{\beta} \,,
\end{align}
where we follow the convention for the Carter tensor in
Boyer-Lindquist coordinates ordered $(t,r,\theta,\varphi)$,
\begin{align}
  \Q^{\alpha\beta} ={}& \diag(-a^{2}\cos^{2}\theta,0,1,\cot^{2}\theta)^{\alpha\beta} \nn\\
  &{}-(a^{2}\cos^{2}\theta) g^{\alpha\beta}
  \,.
\end{align}

Introducing the Mino time parameter $\lambda$ defined
by~\cite{Mino:2003yg},
\begin{align}
  \frac{d\tau}{d\lambda} = \Sigma \,,
\end{align}
the system of ordinary differential equations (ODEs) can be decoupled,
so one would instead integrate the system
\begin{align}
  \label{eq:roflambda}
  \frac{dr}{d\lambda} ={}& \pm_{r} \sqrt{R(r)} \,, \\
  \label{eq:thetaoflambda}
  \frac{d\theta}{d\lambda} ={}& \pm_{\theta} \sqrt{\Theta(\theta)} \,.
\end{align}
Then with solutions for $r(\lambda)$ and $\theta(\lambda)$ in hand,
one can integrate for $\varphi(\lambda)$, $t(\lambda)$, and the
one-to-one function $\tau(\lambda)$ (and thus recover
$x^{\alpha}(\tau)$ if so desired).

The upper/lower signs in Eqs.~\eqref{eq:roflambda} and
\eqref{eq:thetaoflambda} are to be chosen when the particle is
outgoing/ingoing in the radial equation, or downgoing/upgoing in the
polar equation.  A sign flip occurs in an equation when the particle
passes a turning point of the radial or polar motion.  Numerically
integrating this type of equation is inconvenient, as it requires
accurate numerical identification of turning points; and moreover,
when passing through such a turning point, the source in the
differential equation fails to be Lipschitz continuous, becoming
infinitely steep as one approaches the turning point.  Failing the
Lipschitz condition, the Picard-Lindelöf theorem says one can no
longer prove existence and uniqueness of solutions to the ODEs (this
is not a problem for the second order geodesic equations).  Therefore
a reparameterization is necessary.

\subsection{Parameterization for bound orbits}

Hereafter we shall be concerned with bound orbits about a Kerr black hole. 
For such orbits the radial motion is confined within the region $r_p \le
r \le r_a$, where $r_p$ and $r_a$ are the minimum (pericenter) and
maximum (apocenter) radii obtained during the orbital motion, respectively. 
Similarly, the polar motion is confined within the region $\theta_\min \le \theta \le \theta_\max = \pi - \theta_\min$.
An orbit in the equatorial plane has $\theta = \theta_\min = \pi/2$.

There are infinitely many ways to parametrize geodesic motion in Kerr spacetime. 
For bound orbits it is convenient to change from the set
$(\E,\L,\Q)$ to a Keplerian-inspired choice.
One such choice for the radial motion is the quasi-Keplerian
parameterization,
\begin{align}
  \label{eq:r-quasi-kep}
  r=\frac{p M}{1+e\cos\psi}\,,
\end{align}
where $p$ is the dimensionless semi-latus rectum, $0\le e < 1$ is the orbital
eccentricity, and $\psi$ is a monotonically increasing radial phase
parameter.
The minimum (pericenter) and maximum (apocenter) radii occur at
\begin{align}
  \label{eq:r-peri-apo-p-e}
  r_p = \frac{p M}{1+e} \,, \qquad r_a = \frac{p M}{1-e} \,,
\end{align}
which can be inverted to give
\begin{align}\label{eq:p_e_definitions}
	p = \frac{2 r_a r_p}{M (r_a + r_p)} \,, \qquad e = \frac{r_a  - r_p}{r_a + r_p} \,.
\end{align}
Using the parameterization of Eq.~\eqref{eq:r-quasi-kep} will avoid
the issue of sign flipping at turning points, since the radial phase
$\psi$ is monotonically increasing.  Further, one can show that this
parameterization analytically cancels the zeroes in $R(r)$ at $r_{p}$
and $r_{a}$, thus making the ODE satisfy the Lipschitz condition.

A similar approach works for the polar angle.  Defining
$z=\cos\theta$, we can write
\begin{align}
  z = z_{m} \cos\chi
  \,,
\end{align}
where $z_{m}=\cos\theta_{\min}$, so $\pm z_{m}$ are the maximum/minimum values achieved by
$\cos\theta$, and $\chi$ is a monotonically increasing phase angle.
This parameterization similarly solves the sign choice and Lipschitz
continuity issues.

\begin{figure}[tb]
  \begin{center}
\ifdefined\myext
  \tikzsetnextfilename{orbit}
  \tikzstyle{singlearrow} = [draw, arrows={-Latex[length=2mm]}]
\tikzstyle{doublearrow} = [draw, arrows={Latex[length=2mm]-Latex[length=2mm]}]
\tikzstyle{midarrow} = [draw, postaction={decorate}, decoration={markings, mark=at position 0.33 with {\arrow{Latex[length=2mm]}}}]

\begin{tikzpicture}[y=0.80pt, x=0.80pt, yscale=-1.500000,
  xscale=1.500000, inner sep=2pt, outer sep=2pt, auto]
  \path[singlearrow, draw=black,line join=miter,line cap=butt,miter limit=4.00,even odd
    rule,line width=0.800pt] (96.2170,167.6415) -- (96.2170,74.6318)
    node[above] {$\vec{a}$};

  \path[draw=black,miter limit=4.00,line width=0.800pt] (96.2170,167.6415) ellipse
    (2.1083cm and 0.4412cm);

  \path[rotate=-39.99323,draw=black,miter limit=4.00,line width=0.800pt, midarrow]
    (-73.7274,174.4808)arc(191.915:252.114:40.572552 and
    76.487)arc(252.114:312.314:40.572552 and 76.487)arc(-47.686:12.513:40.572552
    and 76.487);

  \path[draw=black,dash pattern=on 0.80pt off 1.60pt,line join=miter,line
    cap=butt,miter limit=4.00,even odd rule,line width=0.800pt] (96.2170,167.6415)
    -- (55.6542,181.0576);

  \path[draw=black,dash pattern=on 0.80pt off 1.60pt,line join=miter,line
    cap=butt,miter limit=4.00,even odd rule,line width=0.800pt] (21.4751,167.6415)
    -- (170.6562,167.6415);

  \path[draw=black,dash pattern=on 0.80pt off 1.60pt,line join=miter,line
    cap=butt,miter limit=4.00,even odd rule,line width=0.800pt] (96.2170,167.6415)
    -- (49.4771,106.2354);

  \path[draw=black,miter limit=4.00,line width=0.800pt, doublearrow]
  (81.4701,148.4937)arc(232.398:296.199:24.168) node[above right]
  {$\theta_{\text{inc}}$} arc(296.199:360.000:24.168);

  \path[draw=black,miter limit=4.00,line width=0.800pt, doublearrow]
    (75.8615,141.2112) arc (232.398:270.000:33.360) node[above left]
    {$\theta_{\min}$};

\end{tikzpicture}
\else
  \ifx\relsstandalone\undefined
    \tikzstyle{singlearrow} = [draw, arrows={-Latex[length=2mm]}]
\tikzstyle{doublearrow} = [draw, arrows={Latex[length=2mm]-Latex[length=2mm]}]
\tikzstyle{midarrow} = [draw, postaction={decorate}, decoration={markings, mark=at position 0.33 with {\arrow{Latex[length=2mm]}}}]

\begin{tikzpicture}[y=0.80pt, x=0.80pt, yscale=-1.500000,
  xscale=1.500000, inner sep=2pt, outer sep=2pt, auto]
  \path[singlearrow, draw=black,line join=miter,line cap=butt,miter limit=4.00,even odd
    rule,line width=0.800pt] (96.2170,167.6415) -- (96.2170,74.6318)
    node[above] {$\vec{a}$};

  \path[draw=black,miter limit=4.00,line width=0.800pt] (96.2170,167.6415) ellipse
    (2.1083cm and 0.4412cm);

  \path[rotate=-39.99323,draw=black,miter limit=4.00,line width=0.800pt, midarrow]
    (-73.7274,174.4808)arc(191.915:252.114:40.572552 and
    76.487)arc(252.114:312.314:40.572552 and 76.487)arc(-47.686:12.513:40.572552
    and 76.487);

  \path[draw=black,dash pattern=on 0.80pt off 1.60pt,line join=miter,line
    cap=butt,miter limit=4.00,even odd rule,line width=0.800pt] (96.2170,167.6415)
    -- (55.6542,181.0576);

  \path[draw=black,dash pattern=on 0.80pt off 1.60pt,line join=miter,line
    cap=butt,miter limit=4.00,even odd rule,line width=0.800pt] (21.4751,167.6415)
    -- (170.6562,167.6415);

  \path[draw=black,dash pattern=on 0.80pt off 1.60pt,line join=miter,line
    cap=butt,miter limit=4.00,even odd rule,line width=0.800pt] (96.2170,167.6415)
    -- (49.4771,106.2354);

  \path[draw=black,miter limit=4.00,line width=0.800pt, doublearrow]
  (81.4701,148.4937)arc(232.398:296.199:24.168) node[above right]
  {$\theta_{\text{inc}}$} arc(296.199:360.000:24.168);

  \path[draw=black,miter limit=4.00,line width=0.800pt, doublearrow]
    (75.8615,141.2112) arc (232.398:270.000:33.360) node[above left]
    {$\theta_{\min}$};

\end{tikzpicture}
  \else
    \includegraphics[width=0.73\columnwidth]{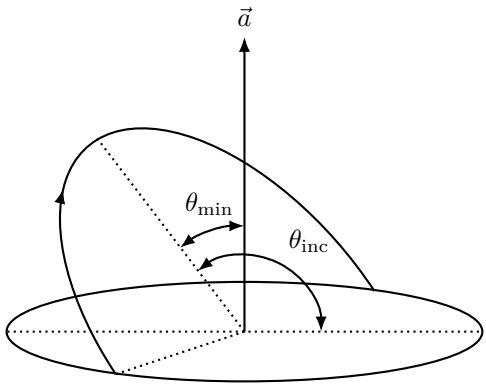}
  \fi
\fi
  \end{center}
  \caption{%
    We parameterize orbital inclination by
    $x=\cos\theta_{\text{inc}}$, see Eq.~\eqref{eq:x-def}.  For
    a prograde orbit, $\theta_{\text{inc}}+\theta_{\min}=\pi/2$,
    whereas for a retrograde orbit (shown here),
    $\theta_{\text{inc}}-\theta_{\min} = \pi/2$.  Using $x$ lifts the
    degeneracy that a single value of $\theta_{\min}$ maps to both
    prograde and retrograde orbits.
  }
  \label{fig:orbit}
\end{figure}

One drawback of using $z_{m}$ as an ``inclination'' parameter is that
it does not distinguish between prograde and retrograde orbits.
This distinction must be implemented by making $a<0$ for retrograde
orbits.  Besides $z_{m}$, there are many common parameterizations for
the inclination angle in the literature.  Because of the plethora of
inclination parameterizations in the literature care must be taken
when comparing results between different works.
In this work we primarily use
\begin{equation}
  \label{eq:x-def}
	x = \sin[\sign(\L)\theta_\min] = \cos\theta_{\text{inc}} \,.
\end{equation}
This has the nice property that the orbital parameters smoothly vary
from prograde equatorial motion ($x=1$) to retrograde equatorial
motion ($x=-1$), without having to flip the sign of $a$.
The relationship between $\theta_{\min}$ and $\theta_{\text{inc}}$ is
diagrammed in Fig.~\ref{fig:orbit}.
The parameters $x$ and $z_{m}$ satisfy the polynomial relationship
$x^{2}+z_{m}^{2}=1$, which is significant in that any polynomial
results developed with $x^{2}$ will also be polynomial in $z_{m}^{2}$.
Another commonly used inclination angle is $\cos\iota = \L/\sqrt{\L^2
  + \Q}$.  Using $\cos\iota$ or $\sin\iota$ also turns out to yield
polynomial relations below.

When parameterizing an orbit by $(p,e,x)$, it is crucial to know how
to convert back and forth between these parameters and the constants
$(\E,\L,\Q)$.  The bijective relationship \cite{Warburton:2013yj} between
$(p,e,x)\leftrightarrow(\E,\L,\Q)$ is well known for bound orbits
in Schwarzschild spacetime \cite{Cutler:1994pb} as well as equatorial
orbits \cite{Glampedakis:2002ya}, spherical orbits
\cite{Hughes:1999bq}, and generic orbits in Kerr spacetime
\cite{Schmidt:2002qk}.
Unfortunately, not all of $(p,e,x)$ space maps to stable bound
orbits, or even to physically realizable motion.  Finding the
separatrix between the stable and plunging orbits is the subject of
the remainder of this paper.

\subsection{Orbit naming conventions}

Certain classes of orbital configurations are simpler to analyze than
others. These special classes of orbits are as follows:
\begin{enumerate}
	\item Equatorial orbits. These lie in the equatorial plane ($\theta = \pi/2$) and have $|x|=1$.
	\item Polar orbits. These orbits have $x = 0$ which corresponds to $\L=0$. They intersect the axis of symmetry of the black hole.
	\item Spherical orbits. These orbits have $e=0$ and ${|x| \neq 1}$. These orbits have fixed Boyer-Lindquist radius and librate in the polar direction.
	\item Circular orbits. These orbits have $e=0$ and $|x| = 1$. These orbits lie in the equatorial plane and have a fixed Boyer-Lindquist radius.
  \item Parabolic trajectories.  These have $e=1$ and ${\E=1}$, sending
    apocenter to infinity, so are marginally bound and technically not
    orbits.
\end{enumerate}
If an orbital configuration does not fall into any of the above categories we refer to it as a `generic' orbit.

\section{The separatrix and\\* other special orbits}\label{sec:separatrix}

The separatrix is the locus of points in the $(p,e,x)$ parameter space 
which separates bound orbital motion from trajectories that plunge
into the black hole.
At fixed $a$,
the separatrix forms a two dimensional surface bounded within $0 \le e
\le 1$ and $-1 \le x \le 1$ in $(p,e,x)$ space.
For parameters in this range we define the location of the separatrix as $\psep(a,e,x)$.

In the literature, orbits with parameters along the separatrix are
referred to as last stable orbits (LSOs) or marginally stable
orbits~\cite{Bardeen:1972fi}.
Orbits along the separatrix with $e=0$ are
referred to as the innermost stable spherical orbit (ISSO). If $|x|=1$
this orbit is usually called the innermost stable circular orbit
(ISCO) instead. At the other extreme there are parabolic orbits with $e=1$.
These orbits have $\E=1$ and are marginally bound.

As the orbital parameters approach the separatrix the amount of azimuthal
precession diverges \cite{Cutler:1994pb, Glampedakis:2002ya}. 
This gives rise to the `zoom-whirl' behavior of orbits near the separatrix \cite{Glampedakis:2002ya}.
In the limit the whirl phase becomes infinitely long and there is a mapping between a spherical orbit at the whirl radius and
the separatrix parameters -- see Ref.~\cite{Levin:2008yp} for the equatorial case and Appendix \ref{apdx:homoclinic_method} for the extension to generic orbits.
Through this relation the marginally bound orbits are related to spherical orbits with $\E=1$.
These orbits are called the innermost bound spherical orbit (IBSO), or if in
the equatorial plane, the innermost bound circular orbit (IBCO).
The majority of this work
is about the separatrix in general but we give additional results for
the IBSO in Appendix~\ref{sec:ibso}.

For the remainder of this work we set $M=1$ for the sake of brevity.

\subsection{The separatrix polynomial}
\label{sec:sep-poly}

Bound radial motion occurs between two roots of the
radial polynomial $R(r)$.  The four roots are traditionally labeled as
$r_{1}\ge r_{2} \ge r_{3} \ge r_{4}$ (when they are all real), in the
factorization
\begin{align}
  \label{eq:R_factorization}
  R(r) = (1-\E^{2}) (r_{1}-r)(r-r_{2})(r-r_{3})(r-r_{4})
  \,.
\end{align}
The signs above are chosen since bound motion happens in the range
$r_{p}=r_{2} \le r \le r_{1} = r_{a}$, and $\E^{2}< 1$ for bound
motion.

When $r_{2} > r_{3}$, there is a simple root, $R(r_{2})=0$ and
$R'(r_{2})\neq 0$, and thus a `restoring force' to keep the particle
from plunging.  By contrast, if we have a root with higher
multiplicity, $r_{2}=r_{3}$ (or $r_{1}=r_{2}=r_{3}$ for circular
orbits), then the derivative of the radial polynomial vanishes,
$R'(r_{2})=0$.  This means there is no `restoring force' at
pericenter, so an infinitesimal perturbation can make the orbit
plunge.

This gives the condition for the separatrix in parameter space:
the set of parameters where these roots degenerate, solving the
equation $r_{2}(p,e,x) = r_{3}(p,e,x)$.

The root $r_{2}=pM/(1+e)$ is a simple function of $p$ and $e$.  The
root $r_{3}$ is much more complicated, though it is possible to
express it in terms of nested radicals (this earlier method is
described in Sec.~\ref{sec:earlier}).
We however pursue an approach which yields the
\emph{separatrix polynomial} $S(a,p,e,x)$, where the separatrix lies
along roots of the polynomial equation $0 = S(a,p,e,x)$.

To find the separatrix polynomial, we start by posing the location of
the separatrix as the simultaneous solutions of the following
system of equations:
\begin{align}
  \label{eq:sys1}
  \begin{cases}
    0 = \Theta(z_{m}) \\
    0 = R(\frac{p}{1-e}) \\
    0 = R(\frac{p}{1+e}) \\
    0 = R'(\frac{p}{1+e})
    \,.
  \end{cases}
\end{align}
It is important to note here that every equation in
system~\eqref{eq:sys1} is a rational polynomial in all the following
indeterminates: $(a,p,e,z_{m},\E,\L,\Q)$.  Since $\Theta$ is a
function only of the square $z_{m}^{2}=1-x^{2}$, this system is still
a system of rational polynomials in $x$ instead of $z_{m}$.  The same
property is true if we use $\sin\iota$ or $\cos\iota$ instead of
$z_{m}$.  We overload the symbol $\Theta$ to mean the appropriate
function of each variable, so we write $\Theta(x)$.

\figoverview

Now if we avoid vanishing denominators (the point is said to be in
general position), we can clear denominators to
form a \emph{system of polynomials} in these indeterminates.  This
system is
\begin{align}
  \label{eq:sys2}
  \begin{cases}
    0 = x^{2}\Theta(x) \\
    0 = (1-e)^{4}R(\frac{p}{1-e}) \\
    0 = (1+e)^{4}R(\frac{p}{1+e}) \\
    0 = (1+e)^{3} R'(\frac{p}{1+e})
    \,.
  \end{cases}
\end{align}
The simultaneous solution of a system of polynomial equations forms an
algebraic variety, the fundamental object of algebraic geometry.  Even
before modern algebraic geometry, in classical elimination theory it
was known that one could eliminate indeterminates from such a system
at the expense of raising the polynomial degree of the remaining
system.  The classic method is based on generalizations and
improvements to Dixon's resultant~\cite{dixon1909eliminant} (see
e.g.~\cite{kapur1994algebraic}).  The more modern algebraic geometry
approach would construct a Gröbner basis for the ideal of the ring of
polynomials vanishing on the variety defined by
system~\eqref{eq:sys2}.

The upshot is that with these four equations, we can eliminate the
three indeterminates $(\E,\L,\Q)$, and be left with the single
\emph{separatrix polynomial} $S(a,p,e,x)$.
As noted earlier, if one desires, the same
approach works in terms of $z_{m}, \cos\iota$, or $\sin\iota$ instead
of $x$.  This can be accomplished with a computer algebra system such
as \textsc{Mathematica} with a bit of guidance.
The separatrix polynomial is 12$^\text{th}$ degree in $p, a, e$ and
4$^\text{th}$ degree in $x^{2}$, so is a bit cumbersome.  After
clearing some unwanted denominators introduced by elimination, we
present $S$ as
\begin{align}
  \label{eq:S-coeffs}
  S = \sum_{n=0}^{12} S_{n} p^{n} \,,
\end{align}
where the $S_{n}$ are polynomials in $a$, $e$, and $x^{2}$ that we
tabulate in App.~\ref{sec:explicit-s}.
We make the algebraic derivation and machine-readable expressions
available in the companion \textsc{Mathematica} notebook distributed with this
article~\cite{SepSupplement}.

Figure~\ref{fig:overview} gives an overview of the set of points
satisfying $S=0$.  The polynomial $S$ is even in $x$, which is
reflected in the reflection symmetry in the figure.  Correspondingly,
the polynomial is satisfied at both the prograde and retrograde values
of $p$ associated to a particular $x^{2}$.

The separatrix itself is just one of the ``leaves'' of the solutions
seen in Fig.~\ref{fig:overview}, specifically the one for which $p$
decreases with increasing $x$, as prograde orbits exist closer than
retrograde ones.  Note also that there are additional unphysical
solutions which appear at smaller $p$ at high spin and eccentricity --
more on this in Sec.~\ref{sec:number-roots}.

\begin{figure}[tb]
  \begin{center}
    \includegraphics[width=\columnwidth]{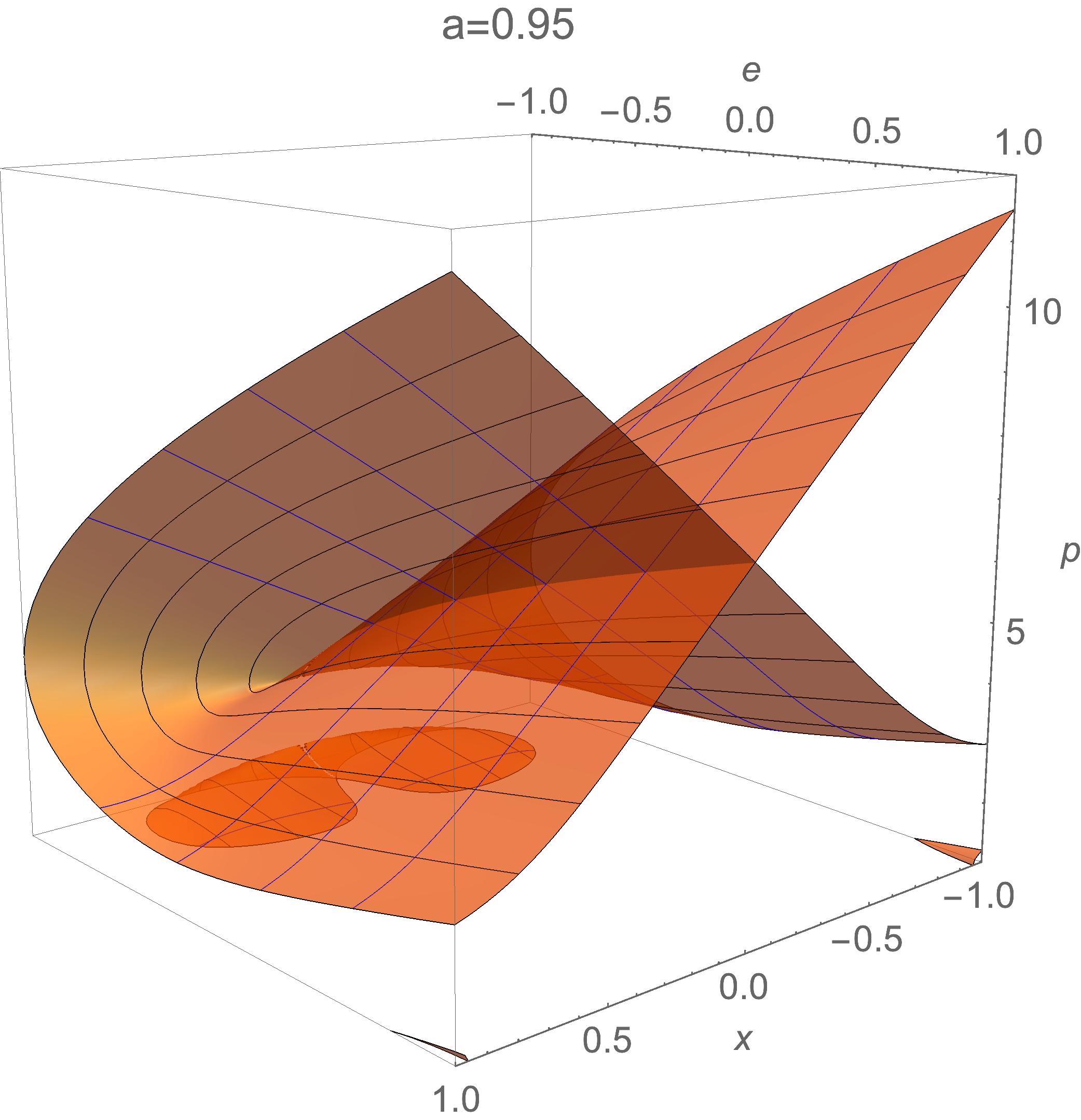}
  \end{center}
  \caption{%
    Extended view, past the physical region, to show that the
    prograde/retrograde branches smoothly join at $e=-1$.  Since
    they are part of the same surface, there is no possibility of a
    factorization for $S$ to give a lower-degree polynomial to describe
    only one of the two branches.
  }
  \label{fig:zoomed-out}
\end{figure}

One might hope that the separatrix polynomial could be factorized into
a lower-degree polynomial for each ``leaf'' of the solution set.
Taking a more global view shows that this is impossible.  In
Fig.~\ref{fig:zoomed-out} we show a view of the affine variety
(i.e.~the set of solutions) that extends to unphysical values of
$e<0$.  There we see that at $e=-1$, the prograde and retrograde
``leaves'' are smoothly connected, so they are part of the same
algebraic set.

The high degree of the separatrix polynomial makes finding analytic solutions for generic orbits unfeasible. Instead we now concentrate on analytic results for interesting limiting cases. We then present techniques for numerically computing the location of the separatrix for generic orbits.

\section{Limiting cases}
\label{sec:limits}

The complete information about the separatrix is contained in the
single polynomial $S(a,p,e,x)$.  This makes it very expedient to take
various simplifying limits, wherein the degree of the polynomial
reduces and thus simplifies.

\subsection{Schwarzschild}
\label{sec:schwarzschild}

As Schwarzschild spacetime is spherically symmetric, the separatrix must be
independent of $x$.  If we set $a=0$, we get the enormous simplification,
\begin{align}
  \label{eq:S-schw}
  S(a=0) = p^{10}(p-6-2e)^{2} \,.
\end{align}
Thus we see that the Schwarzschild separatrix lies at
$\psep[\text{Schw}] = 6+2e$ \cite{Cutler:1994pb}.

\subsection{Equatorial}
\label{sec:equatorial}

Equatorial orbits have $x^{2}=1$, with the sign encoding
prograde/retrograde motion.  This corresponds to $\Q=0$ and
$\sin\iota=0$.  In this case we also get a significant factorization,
\begin{align}
  S(x^{2}=1) = p^{8} S_{\equat}(a,p,e) \,,
\end{align}
where the nontrivial quartic polynomial is
\begin{align}
  \label{eq:S-equat}
  S_{\equat}(a,p,e) ={}& a^4 (-3 - 2 e + e^2)^2 \\
  & {}+ p^2 (-6 - 2 e + p)^2 \nn \\
  & {}- 2 a^2 (1 + e) p (14 + 2 e^2 + 3 p - e p)
  \,.\nn
\end{align}
This is the same as Eq.~(B7) in \cite{OShaughnessy:2002tbu} which is itself a simplified form of Eq.~(23) in \cite{Glampedakis:2002ya}. 
As Eq.~\eqref{eq:S-equat} is quartic in $p$, there is an explicit solution by radicals for $\psep[\pm\equat](a,e)$.

Because of monotonicity, the global extrema are at $a=1, x=\pm 1$.
Along the extremal spin limit $a=1$, the quartic is
\begin{align}
  \label{eq:sep-equat-extremal}
  S_\equat(1,p,e) ={}& (1 + e - p)^2 \\
  &{}\times
  \bigl((-3 + e)^2 - 2 (5 + e) p + p^2\bigr) \nn\,.
\end{align}
The global minimum is at $a=1, e=0, x=+1$, with value
$\psep[\min] = 1$.  The global maximum is at $a=1, e=1, x=-1$, with
value $\psep[\max] = 6 + 4 \sqrt{2} \approx 11.66$.
These two extremes bracket all values for physical solutions of the
separatrix,
\begin{align}
  1 \le \psep(a,e,x) \le 6 + 4 \sqrt{2}
  \,.
\end{align}

In the equatorial plane the separatrix is a one dimensional curve in the $(p,e)$ parameter space.
A parameterization for this curve can be found by noting the connection between the separatrix and the unstable `whirl' radius, $r_u = p/(1+e)$ of a homoclinic orbit \cite{Levin:2008yp}.
This radius varies in the range $r_\ibco \le r_u  \le r_\isco$.
The second inequality implies that $r_u(r_u - 2)a + a^3 \ge 0$.
Simultaneously solving the set of equations $\{S_\equat = 0, r_u = p/(1+e)\}$ with the above constraint gives
\begin{align}
	\esep 	&= \frac{-r_u^2 + 6 r_u - 8 a r_u^{1/2} + 3a^2}{r_u^2 - 2r_u +a^2},	\\
	\psep 	&= \frac{4 r_u (r_u^{1/2} - a)^2}{r_u^2 - 2 r_u + a^2}.
\end{align}
These equations agree with the results in Ref.~\cite{Levin:2008yp}.

\subsubsection{Number of real equatorial roots and brackets}
\label{sec:number-roots}

\begin{figure}[tb]
  \centering
  \includegraphics[width=\columnwidth]{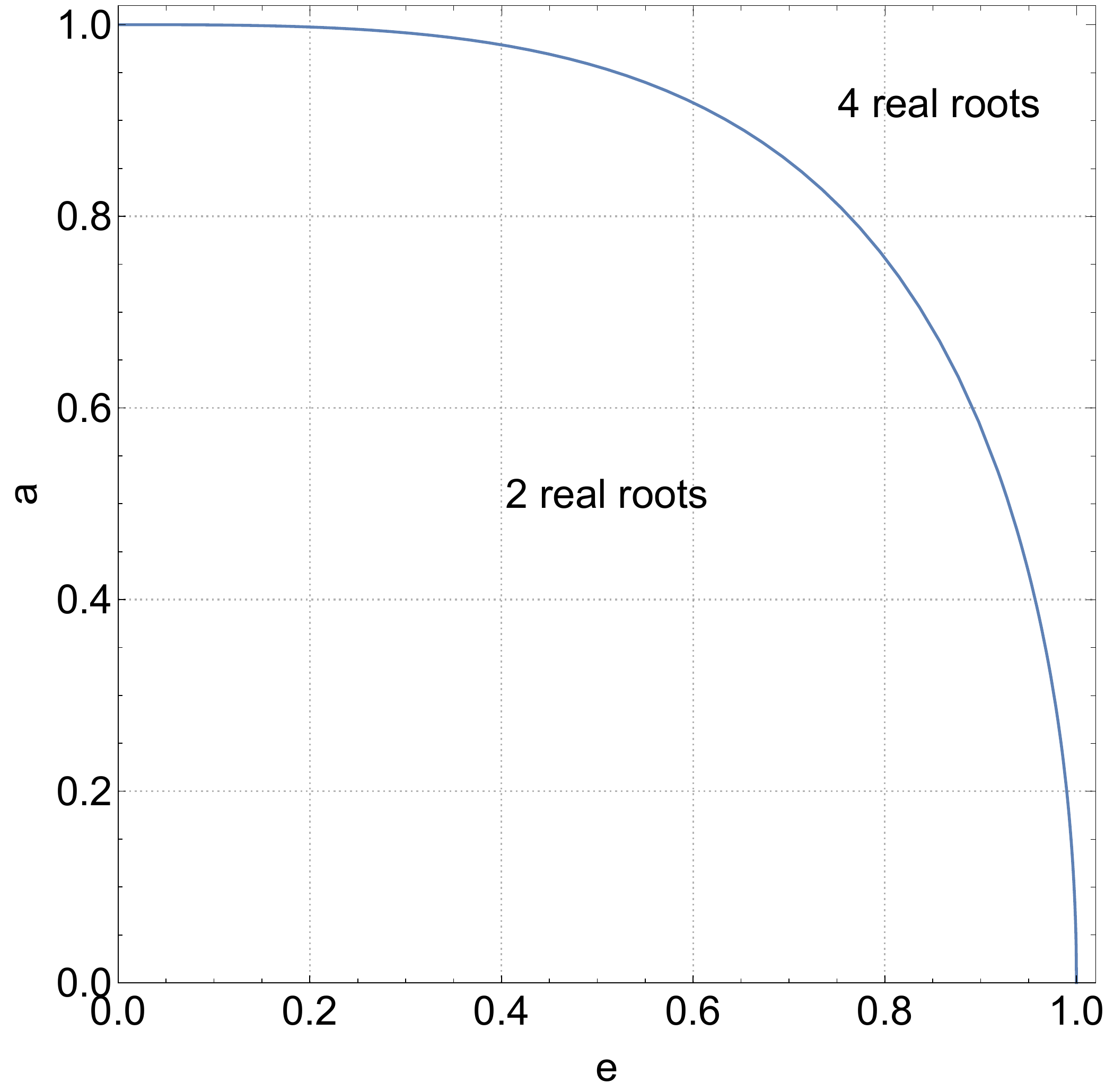}
  \caption{%
    Number of real roots of the equatorial separatrix
    polynomial $S_\equat(p)$.  Along the curve
    $a_{\text{dgn.}}(e)$ given by Eq.~\eqref{eq:a-degen}, there are 4
    real roots but only 3 distinct values, as one pair of complex
    conjugate roots have degenerated and become real.}
  \label{fig:equat-degen-plot}
\end{figure}

The quartic $S_{\equat}(p)$ has four roots, but they are not all
real throughout the unit square in $(a,e)$ space.  We can find the
positions of degenerate roots in parameter space by examining the
discriminant of $S_{\equat}$ when treated as a polynomial in
$p$.  This discriminant is
\begin{multline}
  \label{eq:equat-disc}
  \Delta_{\equat}(a,e) =
  2^{20}
  a^6  (a^{2}-1) (1 + e)^4\\
  \times\Bigl(a^2 (-3 + e)^3 (1 + e) - (-1 + e) (3 + e)^3\Bigr)
  \,.
\end{multline}
Degeneracies occur when the discriminant vanishes.  At $a=0$, the
prograde and retrograde separatrices coalesce and are thus
degenerate.  The degeneracy at $e=-1$ is unphysical but can be seen in
Fig.~\ref{fig:zoomed-out} as the location where the prograde and
retrograde sheets smoothly connect.  The discriminant's other roots
are at $a^{2}=1$ or when
\begin{align}
  \label{eq:a-degen}
  \left(a_{\text{dgn.}}(e)\right)^{2} = \frac{(1-e)(3+e)^{3}}{(1+e)(3-e)^{3}}
  \,.
\end{align}
That is, $S_{\equat}(p)$ has degenerate roots when
$a=a_{\text{dgn.}}$.  Since $S_{\equat}$ has real coefficients,
its roots are either real or come in complex-conjugate pairs.  For
general values of $(a,e)$, there are either 4 real roots or 2 real and
a complex conjugate pair.  However along $(a_{\text{dgn.}}(e),e)$,
there are 4 real roots but one pair has multiplicity 2.  The root with
multiplicity 2 has value
\begin{align}
  \label{eq:p-mul}
  p_{\text{mul}} = \frac{3-2e-e^{2}}{3-e}
  \,,
\end{align}
and note that $0 \le p_{\text{mul}} \le 1$, with equality occurring at
the endpoints in $e$.  Since this is less than or equal to 1, it is
always unphysical, except at $(a=1,e=0)$.  The number of equatorial
roots as a function of $(a,e)$ is summarized in
Fig.~\ref{fig:equat-degen-plot}.

Since $p_{\text{mul}}$ appears at an unphysically small value of $p$,
we can safely ignore these extra solutions, and focus on the outermost
two, which are, in increasing magnitude, the prograde and retrograde
equatorial separatrices $\psep[\pm\equat](a,e)$, respectively.
Again because of monotonicity, we can bracket the location of these
roots by looking at the Schwarzschild and extremal limits.  As before
for $a=0$, $\psep=6+2e$.  The $x=\pm 1, a=1$ polynomial was previously
given in factorized form in Eq.~\eqref{eq:sep-equat-extremal}.  The
roots are all real, and they are (ordered by value)
\begin{align}
  p &= 5+e-4\sqrt{1+e} \,, \\
  p &=1+e \quad \text{(twice)} \,,\\
  p &= 5+e+4\sqrt{1+e} \,.
\end{align}
The smallest of these is less than 1 and hence one of the unphysical
solutions.  The largest of these is the retrograde, equatorial,
extremal separatrix.  The intermediate value is the prograde,
equatorial, extremal separatrix \cite{Glampedakis:2002ya}, and one of the unphysical roots has
degenerated with this physical one.

These extremal ($a=1$) values of the separatrix give us brackets for
general values of the equatorial separatrix,
\begin{align}
  1+e &\le \psep[+\equat](a,e) \le 6+2e \,,\\
  6+2e &\le \psep[-\equat](a,e) \le 5+e+4\sqrt{1+e} \,
  \,.
\end{align}

\subsection{Polar}
\label{sec:polar}

Polar orbits have $x=0$, $\sin\iota =1$, which corresponds to $\L=0$.
Recall that the full separatrix polynomial $S(p,e,x)$ is only a
function of $x^{2}$ and thus even in $x$.  We also know that the
function $\psep(x)$ is monotonic in $x$, with $\psep$ being smaller for
prograde (positive) and larger for retrograde (negative) values of
$x$.  Therefore as $x$ goes through 0, the physically relevant sheet
of the solution set is crossing through the polar value
$\psep[\text{pol}]$, in a simple root.  But since $S$ depends only on
$x^{2}$, the unphysical sheet is simply the reflection of the physical
one with $x\to -x$.  Thus there is a degeneracy at $x=0$, and further
the polynomial factors as a square of a sextic.  That is,
\begin{align}
  S(x=0) = S_{\text{pol}}(a,p,e)^{2} \,,
\end{align}
where the sextic is
\begin{align}
  \label{eq:S-polar}
  &S_{\text{pol}}(a,p,e)= p^5 (-6 - 2 e + p) \\
  &\quad + a^2 p^3 (-4 [-1 + e] [1 + e]^2 + [3 + e (2 + 3 e)] p) \nn\\
  &\quad- a^4 (1 + e)^2 p (6 + 2 e^3 + 2 e [-1 + p] - 3 p - 3 e^2 [2 + p]) \nn\\
  &\quad+ a^6 (-1 + e)^2 (1 + e)^4 \,.\nn
\end{align}

From monotonicity, the minima and maxima of the polar separatrix occur
respectively at $(a=1,e=0)$ and $(a=0,e=1)$.  That is, for arbitrary
$a$ and $e$, the polar separatrix always lies in this interval,
\begin{align}
\psep[\text{pol}](1,0) \le \psep[\text{pol}](a,e) \le \psep[\text{pol}](0,1) \,,
\end{align}
with equality only at the appropriate corners of the unit square in
$(a,e)$ space.

When $a=0$, we have to recover Eq.~\eqref{eq:S-schw}, and indeed here
we get $S_{\text{pol}}(a=0) = p^{5}(p-6-2e)$, again leading to
$\psep[\text{Schw}]=6+2e$.  Thus the global maximum for the polar
separatrix is $\psep[\text{pol}](0,1)=8$.

For the extremal limit $a=1$ there is no major simplification.  But at
the endpoints $e=0$ and $e=1$ there is, respectively,
\begin{align}
  \label{eq:S-polar-extremal}
  S_{\text{pol}}(a=1,e=0) &= (-1 + p)^2 (1 - 4 p - 6 p^2 - 4 p^3 + p^4) \\
  S_{\text{pol}}(a=1,e=1) &= p^2 (16 + 8 p^2 - 8 p^3 + p^4) \,.
\end{align}
From these we can find the real nontrivial roots of the separatrix
polynomial at these corners of the unit square,
\begin{align}
  \psep[\text{pol}](a=1,e=0) ={}& 1 + \sqrt{3} + \sqrt{3 + 2 \sqrt{3}}\,, \\
  \approx{}& 5.27 \,,\nn\\
  \psep[\text{pol}](a=1,e=1) ={}& \tfrac{2}{3}  \Bigl(3 + (54 -6 \sqrt{33})^{1/3} \nn \\
  &\quad+ \bigl(6 (9 + \sqrt{33})\bigr)^{1/3}\Bigr)\,,\\
  \approx{}& 6.77 \,.\nn
\end{align}
Thus the global minimum for the polar
separatrix is $\psep[\text{pol}](1,0)\approx 5.27$.

\subsection{Circular and spherical orbits}
\label{sec:circ}

Specializing to spherical orbits entails setting $e=0$.
Unfortunately the separatrix polynomial does not factor further at
$e=0$.  This polynomial, $S(e=0)$, agrees with one previously
presented in Appendix~A of Ref.~\cite{Stone:2012tr}, based on the
homoclinic orbit approach of Ref.~\cite{Levin:2008yp}.

Of course we can take further special cases where it does
factor.  For example, the equatorial circular separatrix polynomial is
\begin{align}
  S(e=0,x=1) = p^{8}
  \Bigl(
  9 a^4 + (-6 + p)^2 p^2 - 2 a^2 p (14 + 3 p)
  \Bigr)
  \,.
\end{align}
The quartic can be solved by radicals, and the two physical solutions
(prograde and retrograde) are the classic result of the equatorial
ISCO given by Bardeen, Press, and Teukolsky~\cite{Bardeen:1972fi}.

Another special case of interest is spherical orbits around an extremal
black hole.  Here we get the factorization
\begin{align}
  \label{eq:S-circ-extr-full}
  S(e=0,a=1) = (p-1)^{3} S_{\text{sph., ext.}}(p,x)
  \,,
\end{align}
where the ninth degree factor is
\begin{align}
  \label{eq:S-circ-extr-nonic}
  S_{\text{sph., ext.}} ={}& p^{9 } -9 p^8 + 12 p^7 z_{m}^2 \nn\\
  &{}+ 36 p^6 z_{m}^2
  + 30 p^5 z_{m}^4 \nn\\
  &{}- 30 p^4 z_{m}^4 - 36 p^3 z_{m}^6 \nn\\
  &{}- 12 p^2 z_{m}^6
  + 9 p z_{m}^8 - z_{m}^8
  \,,
\end{align}
where $z_{m}^{2}=1-x^{2}$.  This polynomial may be described as
``anti-reciprocal'' in two indeterminates, in the sense that
$S_{\text{sph., ext.}}(p,z_{m}) = - p^{9}z_{m}^{8}S_{\text{sph., ext.}}(p^{-1},z_{m}^{-1})$.

\begin{figure}[t]
  \centering
  \includegraphics[width=\columnwidth]{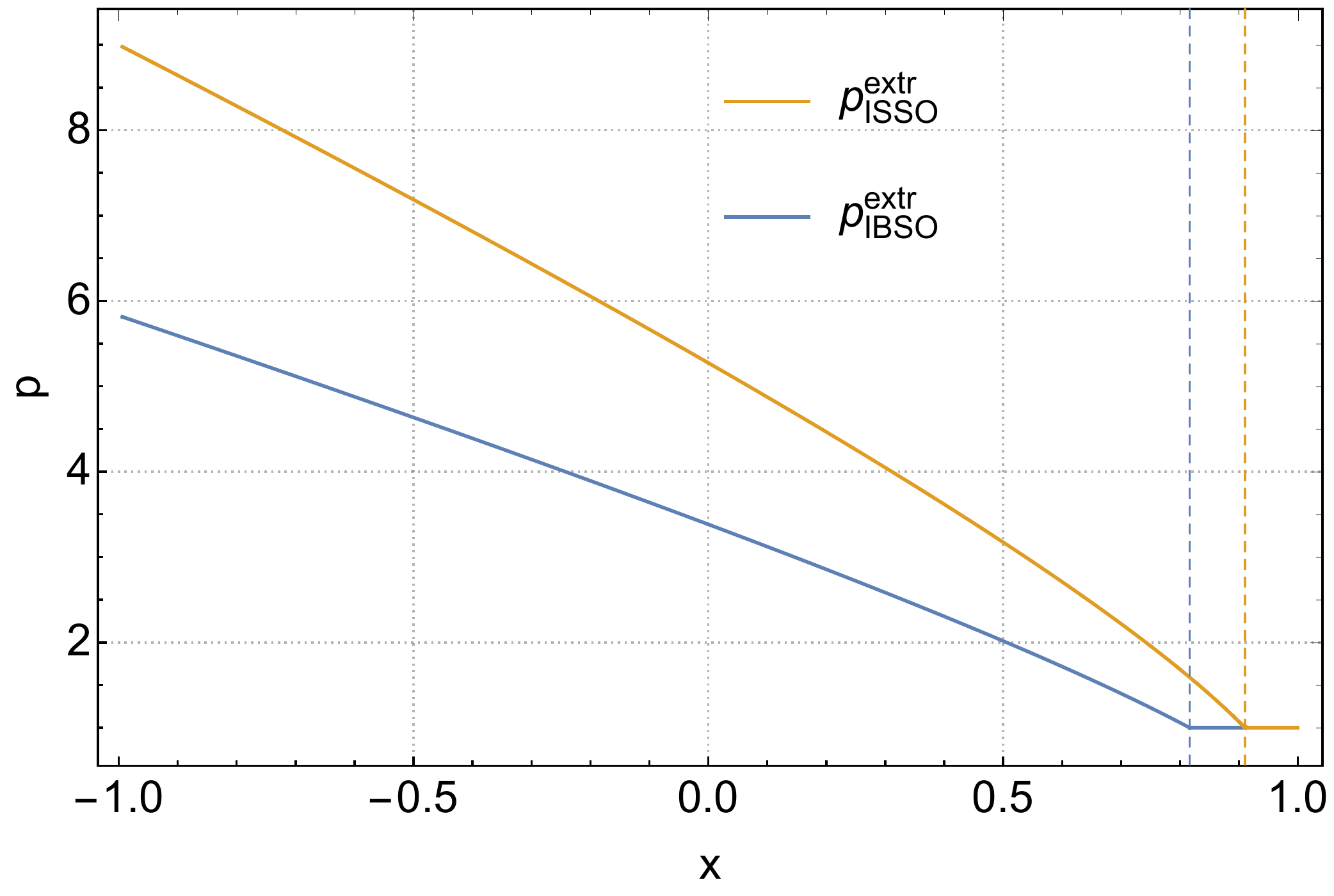}
  \caption{%
	Location of the innermost stable spherical orbit (blue, solid curve) and the innermost bound spherical orbit (yellow, solid curve) about an extremal ($a=1$) black hole.
	Each curve has a kink at a different
    inclination $x$ (marked by the vertical, dashed lines), where two roots of the associated polynomials
    cross each other linearly, as discussed in Secs.~\ref{sec:circ}
    and \ref{sec:extremal-ibso}.
  }
  \label{fig:circ-ibso-extremal}
\end{figure}

Here we see an interesting phenomenon.
For most values of $x$, one root that varies smoothly with $x$ is the
physical root $\psep[\text{sph. ext.}](x)$.  But, this root crosses
$p=1$ linearly at some critical value
$x_{\text{kink}}^{\text{sph. ext.}}$.  Above this value of $x$, the
root $p=1$ becomes the physically relevant root, and this leads to a
kink in the graph of $\psep[\text{sph. ext.}](x)$.
We find the inclination where the nonic also has a root at $p=1$
by setting $p$ to 1, leaving us to solve the polynomial $0 =
8x^{2}(x^{2}-2)(x^{4} + 4x^{2}-4)$.  We find the location of the kink is
\begin{align}
  \label{eq:x-circ-extr-kink}
  x_{\text{kink}}^{\text{sph. ext.}} = \sqrt{2(\sqrt{2} - 1)}
  \,.
\end{align}
Ref.~\cite{Compere:2020eat} independently derived this result at the same time as this work.
From Eq.~\eqref{eq:x-circ-extr-kink} when $a=1$ and $x \ge  x_{\text{kink}}^{\text{sph. ext.}} $, the
ISSO is at $p=1$.  This kink behavior can be seen in Fig.~\ref{fig:circ-ibso-extremal}.

\subsection{Parabolic trajectories}
\label{sec:parabolic-orbits}

Parabolic encounters are astrophysically interesting for modeling
tidal disruptions of ordinary stars around supermassive black
holes~\cite{1989ApJ...346L..13E, Stone:2018nbx, Hayasaki:2018hxv}.
They are also interesting as potential sources of gravitational wave bursts~\cite{Berry:2013poa,Hopper:2017qus}.
A parabolic encounter has $e=1$, which sends the apocenter $r_{1}$ to
infinity while $r_{2} = p/2$ remains finite.  Notice that sending one
root of $R(r)$ to infinity depresses the quartic to a cubic, which
happens when $\E=1$.

Parabolic trajectories are technically not bound orbits. They are
another type of parameter space separatrix, between eccentric (bound)
and hyperbolic (unbound) trajectories.  The set of all parabolic
orbits also connects to the separatrix between bound and plunging
orbits which we are analyzing in this paper, simply by restricting to
$e=1$ in $S(a,p,e,x)$.  This reflects an interesting phase space
geometry which is beyond the scope of this work.

The separatrix polynomial also factorizes at $e=1$,
\begin{align}
  S(e=1) = p^{4} S_{\text{para}}(a,p,x)\,,
\end{align}
where the nontrivial factor is
\begin{align}
  S_{\text{para}}(a,p,x) = {}& 16 a^4 [16 a^4 + 24 a^2 p^2 + p^3 (9 p-32)] x^4 \nn \\
  &{}-8 a^2 [64 a^6 + 80 a^4 p^2 + p^5 (3 p-8) \nn\\
  &\qquad + 4 a^2 p^3 (7 p-24)] x^2 \nn\\
  &{}+[16 a^4 + 8 a^2 p^2 + (p-8) p^3]^2
  \,.
\end{align}
Besides the $p$ degree being lowered, $S_{\text{para}}$ is only
quadratic in $x^{2}$, meaning it is straightforward to give an
explicit parametric description of the surface.
Before doing so, we will further specialize to equatorial parabolic
encounters to find the $p$ extrema of this slice through the
separatrix.  Setting $x=1$, we have the further simplification
\begin{multline}
  S_{\text{para}}(a,p,x=1) = p^4 (4 a^2 - 8 p - 4 a p + p^2) \\
  \times(4 a^2 - 8 p + 4 a p + p^2)
  \,.
\end{multline}
Besides the unphysical roots at $p=0$, there are two additional
quadratic factors which can be solved directly.  The roots of physical
interest give the extrema of the parabolic separatrix as a function of
$a$,
\begin{align}
  p_{\text{para}}^{\pm}(a) = 2(2+2\sqrt{1\pm a}\pm a)\,.
\end{align}
The plus signs are taken for retrograde (larger $p$), and the minus
signs are taken for prograde (smaller $p$).

To find the parametric description of the parabolic separatrix, match
coefficients $(A,B,C)$ in $S_{\text{para}} = A x^{4} + B x^{2} + C$,
then solve for $x^{2}$ in $0=S_{\text{para}}$,
\begin{align}
  \label{eq:para-para}
  x_{\text{para}}^{2}(a,p) = \frac{- B \pm \sqrt{D}}{2A}\,,
\end{align}
where the discriminant is $D=2^{13} a^{4}p^{7}[4 a^2 + p(p-4)]^{2}$.
If one takes the upper sign in Eq.~\eqref{eq:para-para}, the values of
$x_{\text{para}}^{2}$ are always greater than 1 and thus unphysical;
therefore take the minus sign in Eq.~\eqref{eq:para-para}.  The
parameter $p$ lies in the domain
$p_{\text{para}}^{-}(a)\le p \le p_{\text{para}}^{+}(a)$, and
the image $x_{\text{para}}^{2}(a, p)$ covers $[0,1]$ on this domain.

\section{Numerical implementation}
\label{sec:num-impl}

In this section we assume that numerical values are given for
$0\le a\le 1$, $0\le e \le 1$, and $-1 \le x \le +1$.  Then
$S(a,p,e,x)$ is a univariate 12$^{\text{th}}$ degree polynomial in $p$ with real
coefficients, our goal is to find certain real roots.

Specifically, given values for $a$ and $\esep, \xsep$ we present a method to rapidly compute $\psep$. Before we present our new method based upon numerically finding the roots of the separatrix polynomial we review the previous methods in the literature for numerically computing the separatrix.

\subsection{Earlier approaches}
\label{sec:earlier}

Given a set of geodesic parameters, determining if an orbit is stable is straightforward.
Stable orbits have $r_2 - r_3 > 0$, when both are real, and these
roots are easily evaluated as follows.  First, $(\E,\L,\Q)$ can be
expressed explicitly in terms of $(p,e,x)$ by following the procedure
detailed in Appendix B of Schmidt~\cite{Schmidt:2002qk}.  These analytic
expressions involve nested radicals, and thus may become complex.
Next we follow Fujita and Hikida~\cite{Fujita:2009bp} to find $r_{3}$
and $r_{4}$.  Depress the quartic $R(r)$ by the known quadratic factor
$(1-\E^{2})(r_{1}-r)(r-r_{2})$.  This leaves the quadratic
$(r-r_{3})(r-r_{4})$ which is solved explicitly for $r_{3}$ and
$r_{4}$, again in terms of nested radicals, so $r_{3}$ and $r_{4}$ may
also become complex.

\begin{figure}
	\includegraphics[width=8.5cm]{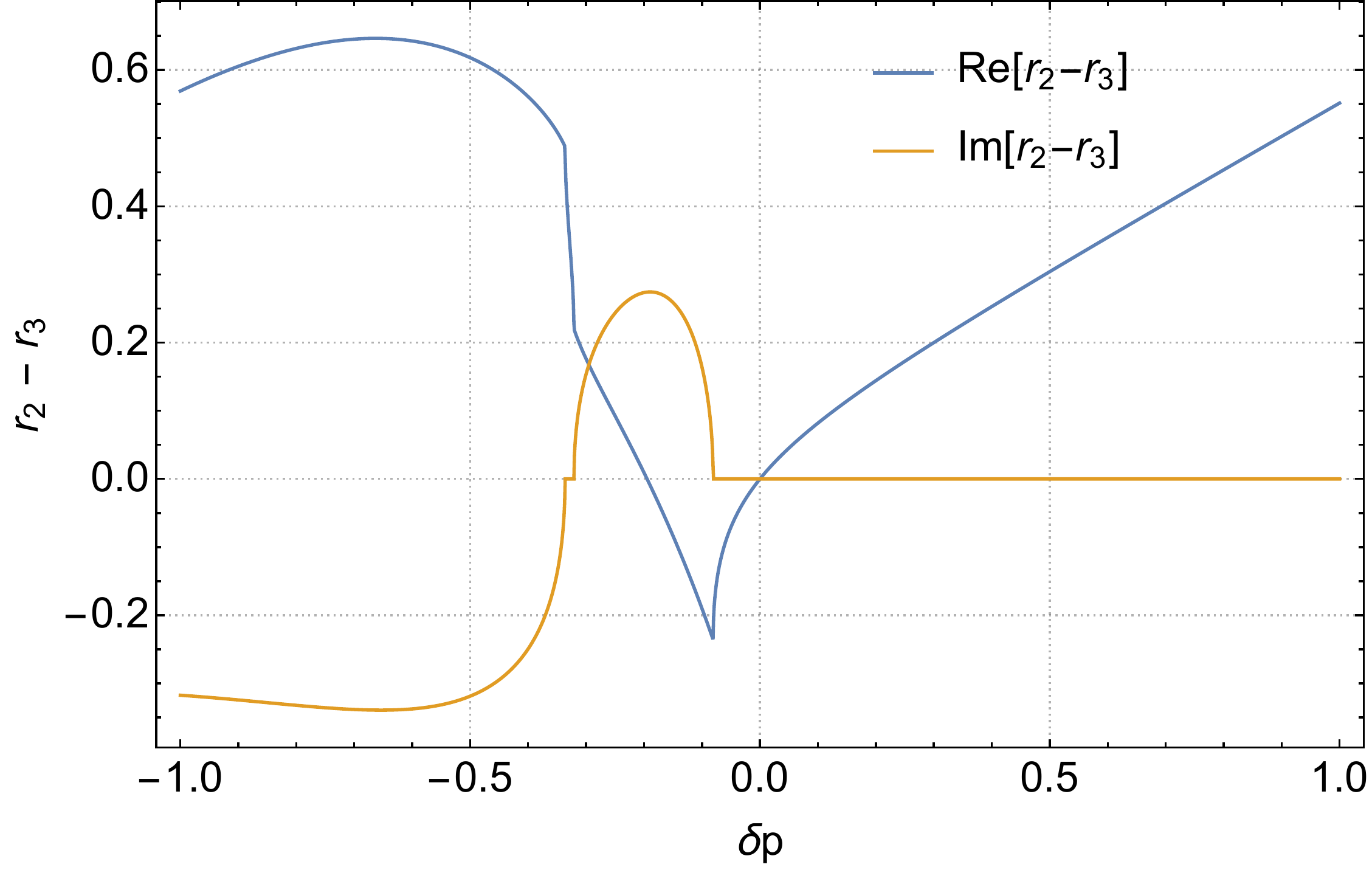}
	\caption{
The difference $r_2-r_3$ plotted as a function of $\delta p = p - \psep$ for $a=0.998, e=0.9, x=0.95$. The plot shows the real and imaginary parts in blue and yellow, respectively. The separatrix is at $\delta p = 0$ which corresponds to $\psep = 2.10085$ where $r_2-r_3 = 0$. The non-smoothness of this function makes it difficult to numerically root find on $r_2-r_3$ to find the separatrix. }\label{fig:r2-r3}
\end{figure}

Finding the precise location of the separatrix in a robust and computationally efficient way is more challenging.
This is because, as mentioned above, in some regions of the parameter space $r_2 - r_3$ becomes complex.
This causes, e.g., a Newton-Raphson root finding scheme to fail -- see Fig.~\ref{fig:r2-r3} and, e.g., Appendix A of \cite{Sundararajan:2008bw}.
These challenging regions of the parameter space occur for high spin, near prograde equatorial orbits.

The problems mentioned above occur when the root finder steps over the separatrix to a value of $p < \psep$ and the algorithm is unsure how to proceed because the function it is evaluating has become complex.
One way to avoid this is to use a bisection-like method to seek the root strictly from above.
In this method you pick an initial value of $p$ large enough to ensure $p > \psep$, check if $r_2 > r_3$ and if so decrease $p$ by some small amount $\Delta \psep$.
This is repeated until $r_3$ becomes complex at which point the previous value of $p$ is returned to and now $\Delta \psep$ is halved and the process repeats.
As this is a bisection-like method it is robust but does not converge quickly.

A robust approach that can use rapidly convergent numerical root finding was introduced into the Black Hole Perturbation Toolkit \cite{BHPToolkit} in 2018 by one of us.
This used an extension of the method of Ref.~\cite{Levin:2008yp} from equatorial to generic orbits.
We give the equations for this extension in Appendix \ref{apdx:homoclinic_method}.
Recently Ref.~\cite{Rana:2019bsn} published equations for a similar extension.
In this approach one picks values for $\esep, \xsep$ and then root finds for an unstable circular orbit radius, $r_u$, in the range $r_\ibso \le r_u $.
With $r_u$ strictly bounded below the method is robust and rapidly convergent numerical root finding techniques, like a Newton-Raphson method, can be employed.
The downside to this method is that first $r_\ibso$ must be found and this adds to the computational overhead.

We now discuss new approaches which are faster and guaranteed to find $\psep$.

\subsection{Global root-finding}
\label{sec:global-root-finding}

From the fundamental theorem of algebra, $S$ has 12 complex roots for
$p$, and several methods exist to find all roots simultaneously.  One
standard ``black-box'' approach~\cite{Press:2007:NRE:1403886} is the
method of Jenkins and Traub.  There exists both a general version for
polynomials with complex coefficients, and a more adapted algorithm
for polynomials with real coefficients~\cite{JenkinsTraub}.

Another popular algorithm is the Aberth method~\cite{MR329236}, which
converges cubically to simple roots, but only exists in a complex
form.  If one were to initiate the Aberth method with purely real
guesses for a real polynomial, the iteration scheme would never push
the guesses into the complex plane, and would thus fail to find
complex roots.  Therefore it is important to start with complex
guesses, though this means that all roots will acquire some imaginary
part, even if the root is exactly on the real axis.  Therefore using
the Aberth method (or the complex version of the Jenkins-Traub method)
requires testing roots $p_{i}$ for realness via
$|\text{Im}[p_{i}]|\le \epsilon$ with some arbitrary choice of
$\epsilon > 0$.

Such black-box global root-finding methods are implemented in most
computer algebra systems and numerical libraries.  For example, in
\textsc{Mathematica}, the command \texttt{NRoots[]} implements both
the Jenkins-Traub and Aberth methods.

For special values such as $x=0$, $x^{2}=1$, or $e=0$, one should use
the appropriate simplified polynomial.  Here we give the generic
algorithm, for general points:
\begin{enumerate}
\item Find all complex roots $p_{i}$, and select the real roots as
  those that satisfy $|\text{Im}[p_{i}]|\le \epsilon$ with some
  $\epsilon > 0$ determined by the required precision.
\item If $x<0$, the orbit is retrograde and thus the desired root is
  the largest real root.
\item If $x>0$, the orbit is prograde and thus the desired root is
  the second largest real root.
\end{enumerate}

\subsection{Real root isolation}
\label{sec:real-root-isol}

For polynomials with real coefficients, it is possible to bound the
number of real roots and to isolate each real root into an interval of
the real line, with black-box ``real-root isolation'' algorithms.  At
their most basic, these algorithms arise from Descartes' rule of
signs~\cite{descartes:1637}, with improvement due to Sturm's
theorem~\cite{basu2013algorithms}.  Using real-root isolation, one is
guaranteed to find brackets for all the simple real roots of $S(p)$
automatically.  Such algorithms are implemented in several computer
algebra systems, for example in the \textsc{sage} module
\texttt{sage.rings.polynomial.real\_roots}, or in the
\textsc{Mathematica} commands \texttt{RootIntervals[]} or
\texttt{NSolve[\ldots,Reals]}.

For special values such as $x=0$, $x^{2}=1$, or $e=0$,
the separatrix polynomial is not square-free, and one should instead
focus on solving the nontrivial factor such as $S_{\text{pol}}$.
Here we give the generic algorithm, for general points:
\begin{enumerate}
\item Find isolating intervals for all roots of $S(p)$ in the
  admissible physical range, $1\le p \le 6 + 4 \sqrt{2} \approx 11.66$.
\item If $x<0$, the orbit is retrograde and thus the desired root is
  the largest real root.
\item If $x>0$, the orbit is prograde and thus the desired root is
  the second largest real root.
\end{enumerate}

\subsection{Bracketing of roots}
\label{sec:bracketing-roots}

Rather than relying on a black box algorithm to find isolating
intervals for real roots, we can analytically find them, using all
the limiting cases presented in Sec.~\ref{sec:limits}.  The method we
describe here is also the fastest and most robust, and the one that is
implemented in the Black Hole Perturbation Toolkit~\cite{BHPToolkit}.

First note that we have the brackets,
\begin{align}
  \label{eq:polish-brack-pro}
  \psep[+\equat](a,e) \le {}&\psep[\text{pro}](a,e,x) \le \psep[\text{pol}](a,e) \,, \\
  \label{eq:polish-brack-ret}
  \psep[\text{pol}](a,e) \le {}&\psep[\text{ret}](a,e,x) \le \psep[-\equat](a,e) \,,
\end{align}
where $\pm$equat refer to the prograde/retrograde equatorial orbits.
These serve as brackets, if we know the values of the polar and
equatorial separatrices.  Those values are found via their own
bracketed root-finding.  The algorithm proceeds as follows, given some
inputs $a,e,x$:
\begin{enumerate}
\item In all cases one needs to find $\psep[\text{pol}](a,e)$, by
  polishing the single simple root of the sextic $S_{\text{pol}}(p)$ given in
  Eq.~\eqref{eq:S-polar}, within the bracket
  \begin{align}
    1 + \sqrt{3} + \sqrt{3 + 2 \sqrt{3}} \le \psep[\text{pol}](a,e) \le 8
    \,.
  \end{align}
\item If $x=0$, the orbit is polar and the separatrix has been found.
  Otherwise:
  \begin{enumerate}
  \item If $x>0$, find $\psep[+\equat](a,e)$, by polishing the
    single simple root of Eq.~\eqref{eq:S-equat} within the bracket
    \begin{align}
      1+e &\le \psep[+\equat](a,e) \le 6+2e \,.
    \end{align}

    Now with $\psep[+\equat](a,e)$ and $\psep[\text{pol}](a,e)$
    in hand, polish the single simple root of the full
    12$^{\text{th}}$ degree $S(p)$ within the bracket
    Eq.~\eqref{eq:polish-brack-pro}.

  \item If $x<0$, one can omit finding $\psep[-\equat](a,e)$,
    since there is only ever one root of the separatrix polynomial
    between $\psep[\text{pol}](a,e)$ and the maximum possible value of
    the separatrix (discussed in Sec.~\ref{sec:equatorial}),
    $\psep[\max] = 6 + 4 \sqrt{2} \approx 11.66$.  Therefore
    polish the single simple root of the full 12$^{\text{th}}$ degree
    $S(p)$ within the bracket $\psep[\text{pol}](a,e) \le \psep(a,e,x)
    \le 12$.
  \end{enumerate}
\end{enumerate}

\subsection{Implementation in the\\* Black Hole Perturbation Toolkit}

The algorithm presented above to compute the separatrix is implemented in the \texttt{KerrGeodesics} \textsc{Mathematica} package of the Black Hole Perturbation Toolkit~\cite{BHPToolkit}.
It can be accessed with a function called \texttt{KerrGeoSeparatrix[a,e,x]}. This algorithm replaced the slower algorithm outlined in Appendix~\ref{apdx:homoclinic_method}. The new method takes \textasciitilde 1ms to calculate the location of the separatrix to machine precision.\footnote{This is measured on a 2.5GHz Macbook Pro laptop using \textsc{Mathematica} 12.} This is roughly 45 times faster than the previous implementation which relied on multiple root finding steps.

In addition to numerically finding $\psep$ the \texttt{KerrGeoSeparatrix[a,e,x]} function will also return the closed form analytic results for the special cases presented in Sec.~\ref{sec:limits}.

\section{Summary and future work}
\label{sec:concl-disc}

In this article we have examined the separatrix between stable bound
orbits and plunging orbits for test body motion in Kerr spacetime. We
found the generic polynomial whose roots are the location of the
separatrix in the $(p,e,x)$ parameter space -- stated in
Eq.~\eqref{eq:S-coeffs}, with coefficients tabulated in
Appendix~\ref{sec:explicit-s} and in the supplementary \textsc{Mathematica}
notebook~\cite{SepSupplement}.  For generic orbits the polynomial is 12$^{\text{th}}$
degree in $p$ and 4$^{\text{th}}$ degree in $x^2$ so in this case
closed form solutions are either not available or practical. For these
orbits we provide robust algorithm for numerically finding separatrix
in Sec.~\ref{sec:num-impl} and provided an example implementation in
the Black Hole Perturbation Toolkit~\cite{BHPToolkit}. For special
classes of orbits the separatrix polynomial simplifies and we can find
analytic solutions. These results are presented in
Sec.~\ref{sec:limits}. In Appendix~\ref{sec:ibso} we also consider the
special $\E=1$ case of the innermost bound spherical orbits (IBSO).

We have focused on bound geodesic motion in the Kerr spacetime
in this work.
Generalizations and extensions are possible.  We expect that in the
Kerr-Newman spacetime [and perhaps even the Kerr-Newman-Taub-NUT-(anti-)de~Sitter
family], the separatrix is also an algebraic variety and can be reduced
to a single polynomial in parameter space.
In a more astrophysically relevant extension, it would
also be interesting to consider the case where the orbiting test body
is spinning. In this case the body's spin couples to the local
curvature of the spacetime~\cite{Mathisson2010,Papa51,Dixo70} and this
modifies the orbital motion~\cite{Ruangsri:2015cvg}.  This in turn
modifies the location of the separatrix.  To the best of our knowledge,
the change to the separatrix due the spin on the test body has only been
studied in the circular, equatorial orbit
case~\cite{Jefremov:2015gza}.

\section*{Acknowledgements}

LCS acknowledges Daniel McLaury for helpful discussion.
NW gratefully acknowledges support from a Royal Society - Science Foundation Ireland University Research Fellowship.

\appendix

\section{Coefficients of $S$}
\label{sec:explicit-s}

This section contains the coefficients $S_{n}$ in the expansion of the
separatrix polynomial
Eq.~\eqref{eq:S-coeffs}, repeated here for convenience,
\begin{align}
S(a,p,e,x) = \sum_{n=0}^{12}S_{n} p^{n}
\,.
\end{align}
As mentioned earlier, it is possible to develop the separatrix
polynomial with the angular parameter being any of
$x, z_{m}, \sin\iota$, or $\cos\iota$.  The relationships
$x^{2}+z_{m}^{2}=1$ and $\sin^{2}\iota + \cos^{2}\iota =1$ allow
converting between pairs of them, so we present coefficients in two
angular parameterizations below.
We have also provided a \textsc{Mathematica} notebook as a machine-readable
supplement to this article, containing the derivation and resulting
polynomial~\cite{SepSupplement}.

\begin{widetext}
\subsection{As a function of $x$}
\label{sec:s-of-x}

\vspace{-1em}

\begin{align*}
S_{12}={}&1\\
S_{11}={}&-4 (3 + e)\\
S_{10}={}&4 (3 + e)^2 + 2 a^2 (3 + 2 e + 3 e^2 - 2 [3 + e (2 + e)] x^2)\\
S_{9}={}&4 a^2 [-7 + e (-7 + e [-13 - 5 e + 4 (3 + e) x^2])]\\
S_{8}={}&-16 a^2 (-1 + e) (1 + e)^2 (3 + e) (-1 + x^2) + a^4 (15 + 20 e + 26 e^2 + 20 e^3 + 15 e^4 - 4 [9 + e (12 + e [18 + e (12 + 5 e)])] x^2 \\*
&{}+ 2 [15 + e (2 + e) (10 + 3 e [2 + e])] x^4)\\
S_{7}={}&-8 a^4 (1 + e)^2 (-1 + x) (1 + x) (-3 + e -  e^2 - 5 e^3 + [15 + e (-5 + 3 e [1 + e])] x^2)\\
S_{6}={}&-4 a^4 (1 + e)^2 (-1 + x) (1 + x) (-2 [11 - 14 e^2 + 3 e^4][-1 + x^2] \\*
&{}+ a^2 [5 + 6 e^2 + 5 e^4 - (5 + e^2 [6 + e (8 + 5 e)]) x^2 + (-1 + e) (3 + e) (3 + e [2 + e]) x^4])\\
S_{5}={}&8 a^6 (-1 + e) (1 + e)^3 (-1 + x^2)^2 (3 + e + e^2 - 5 e^3 + 2 [6 + e (2 + e + e^2)] x^2)\\
S_{4}={}&a^6 (1 + e)^4 (-1 + x^2)^2 (-16 [-3 + e] [-1 + e]^2 [1 + e] [-1 + x^2] \\*
&{}+ a^2 [15 + e (-20 + e [26 + 5 e (-4 + 3 e)]) + 6 x^2 - 2 e (2 + e) (2 + e [-6 + 5 e]) x^2 + (-1 + e)^2 (3 + e)^2 x^4])\\
S_{3}={}&-4 a^8 (-1 + e) (1 + e)^5 (-1 + x^2)^3 (7 - 7 e + 13 e^2 - 5 e^3 + [-1 + e] [7 + e^2] x^2)\\
S_{2}={}&2 a^8 (-1 + e)^2 (1 + e)^6 (-1 + x^2)^3 (2 [-3 + e]^2 [-1 + x^2] + a^2 [-3 + 2 e - 3 e^2 + (-1 + e) (3 + e) x^2])\\
S_{1}={}&-4 a^{10} (-3 + e) (-1 + e)^3 (1 + e)^7 (-1 + x^2)^4\\
S_{0}={}&a^{12} (-1 + e)^4 (1 + e)^8 (-1 + x^2)^4
\end{align*}

\subsection{As a function of $\sin\iota$}
\label{sec:s-of-iota}

Here we use the shorthand $s=\sin\iota$.
\begin{align*}
S_{12}&=1\\
S_{11}&=-4 (3 + e)\\
S_{10}&=4 (3 + e)^2 + 2 a^2 (-3 -2 e + e^2 + 2 [3 + e (2 + e)] s^2)\\
S_{9}&=-4 a^2 (7 + e [7 + e + e^2 + 4 e (3 + e) s^2])\\
S_{8}&=16 a^2 (-1 + e) (1 + e)^2 (3 + e) s^2 + a^4 [(-3 + e)^2 (1 + e)^2 + 2 (1 + e) (-15 + e [-5 + e (-1 + 5 e)]) s^2 + 4 [3 + e (2 + e)]^2 s^4]\\
S_{7}&=8 a^4 (1 + e)^2 s^2 (15 -5 e + e^2 -3 e^3 -2 [9 + e (-3 + e + e^2)] s^2)\\
S_{6}&=4 a^4 (1 + e)^2 s^2 (2 [-1 + e^2] [7 + e^2 + 2 (-9 + e^2) s^2] + a^2 [-4 + 9 s^2 + e^2 (-2 + 2 [-2 + e] e + [8 + e (4 + 3 e)] s^2)])\\
S_{5}&=-8 a^6 (-1 + e) (1 + e)^3 s^2 (3 + e - e^2 + e^3 + 2 [-3 - e + 2 e^3] s^2)\\
S_{4}&=a^6 (1 + e)^4 s^2 (16 [-3 + e] [-1 + e]^2 [1 + e] s^2 + a^2 [2 (-3 + e) (-1 + e)^2 (1 + e) + (21 + e [-28 + e (22 + e [-12 + 13 e])]) s^2])\\
S_{3}&=-4 a^8 (-1 + e) (1 + e)^5 (-7 + e [7 + e (-13 + 5 e)]) s^4\\
S_{2}&=2 a^8 (-1 + e)^2 (1 + e)^6 (2 [-3 + e]^2 + a^2 [3 + e (-2 + 3 e)]) s^4\\
S_{1}&=-4 a^{10} (-3 + e) (-1 + e)^3 (1 + e)^7 s^4\\
S_{0}&=a^{12} (-1 + e)^4 (1 + e)^8 s^4
\end{align*}
\end{widetext}

\section{Series in spin $a$}
\label{sec:series}

Deriving a series solution for $\psep(a,e,x)$ is very straightforward
since we have the explicit polynomial $S(a,p,e,x)$.  The only
difficulty is that while $S$ has multiple solutions, our series
must be able to pick an individual ``leaf.''  This means we have
to make a branch choice at some point.

At $a=0$, we saw in Sec.~\ref{sec:schwarzschild} that the separatrix
is given by $\psep[\text{Schw}]=6+2e$.  Now we pose the ansatz
\begin{align}
  \psep = \sum_{k=0}^{\infty} p_{k}(e,x) a^{k} \,,
\end{align}
where $p_{0}(e,x) = 6+2e$.  This ansatz can be inserted into the
polynomial $S$ and solved order-by-order in $a$.  At linear order
there are two possible solutions, as mentioned before, and we have to
make a branch choice to pick the physical leaf.  If we truncate at
order $a^{2}$, we have to solve
\begin{align}
  0 = 1024 a^{2} (3+e)^{9} [(3+e)p_{1}^{2} - 32 (1+e)x^{2}] +\mathcal{O}(a^{3})\,.
\end{align}
The two solutions for $p_{1}$ are the prograde and retrograde
leaves, which have coalesced in the $a\to0$ limit.  We need $\psep$ to
decrease with increasing $x$, so we choose the sign
\begin{align}
  p_{1}(e,x) = - x \sqrt{\frac{32(1 + e)}{3 + e}} \,.
\end{align}
This first term was previously found for the equatorial case in~\cite{Glampedakis:2002ya}.
After this sign has been fixed, all higher terms $p_{k}$ come from
solving a linear equation by truncating at order $a^{k+1}$.  The first
few of these are
\begin{align*}
  p_{2} ={} & \frac{-11 - e^3 + 4 x^2 + e (-11 + 4 x^2) + e^2 (-9 + 8 x^2)}{2 (3 + e)^2} \\
  p_{3} ={} & \frac{(1 + e)^{1/2} x }{\sqrt{2} (3 + e)^{7/2}} \biggl(-4 [7 + e (7 + 6 e)] \\
  & \qquad + 3 \Bigl(5 + e [5 -  (-7 + e) e]\Bigr) x^2\biggr)
\end{align*}
and so on up to arbitrary order.
In the companion \textsc{Mathematica} notebook~\cite{SepSupplement}, we provide these coefficients
up through and including $p_{6}$.

Unfortunately, this series is not very useful at high spin.
If we keep terms up to $a^{6}$,
the maximum error in $p$ across $(e,x)$ is $\sim 1\%$ when $a=0.8$.
But by a spin of $a=0.95$, the maximum error is already $\sim 10\%$.
Therefore we do not recommend the series approach.

\section{Innermost bound spherical orbits}
\label{sec:ibso}

The location of the innermost bound spherical orbits (IBSOs) can be
found following the same approach as that for finding the separatrix
polynomial.  Bound orbits have $\E < 1$, and the limit $\E \to 1$
gives marginally bound orbits.  Note that marginally bound spherical orbits are
not stable, being interior to the innermost stable circular
orbits~\cite{Bardeen:1972fi}.  They are interesting nonetheless so we
demonstrate how to find the IBSO polynomial.

We can again form a polynomial system to define the location of the
IBSO.
By setting $\E=1$, we depress the quartic to a cubic, but we will
still number the remaining real roots as $r_{1}\ge r_{2} \ge r_{3}$.
Being a spherical orbit, $r_{1}=r_{2}=p$ is a double root.
Further we are not interested in stability, only the existence of the
spherical orbit with $\E=1$.  Thus our system is
\begin{align}
  \label{eq:sys-ibso}
  \begin{cases}
    0 = x^{2}\Theta(x; \E =1) \\
    0 = R(p; \E=1) \\
    0 = R'(p; \E =1) \,.
  \end{cases}
\end{align}
The last two equations implement the condition that $p$ is a double
root and hence spherical.  This system is polynomial in
$(a,p,x,\L,\Q)$.  We can again eliminate $\L,\Q$ with a computer
algebra system leaving a single polynomial in $(a,p,x)$.  After
removing some unimportant prefactors, we have the IBSO polynomial,
\begin{align}
  \label{eq:I-poly}
  I ={}& (-4 + p)^2 p^6 + 2 a^2 p^5 (-8 + 2 p + 4 x^2 - 3 p x^2) \nn\\
  &{}+ a^4 p^3 [-8 (1 - 3 x^2 + 2 x^4) + p (6 - 14 x^2 + 9 x^4)] \nn\\
  &{}+ 2 a^6 p^2 (2 - 5 x^2 + 3 x^4) + a^8 (-1 + x^2)^2
  \,.
\end{align}

As with the separatrix polynomial, we can take limits and get
simplifications.

\subsection{Schwarzschild IBSO}
\label{sec:schw-ibso}

Setting $a=0$ we get the factorization,
\begin{align}
  \label{eq:schw-I}
  I(a=0) = p^{6}(p-4)^{2}
  \,,
\end{align}
recovering the Schwarzschild IBSO at $p=4$.

\subsection{Equatorial IBSO}
\label{sec:equatorial-ibso}

Setting $x=1$ we get the factorization,
\begin{align}
  \label{eq:equat-IBSO}
  I(x=1) = p^{4}(p^{2} - 4 p - 2 a p + a^2) (p^2 - 4 p + 2 a p + a^2)
  \,.
\end{align}
The quadratic factors can be solved by radicals.  Two of the roots are
in the physical region, recovering the classical result
from~\cite{Bardeen:1972fi},
\begin{align}
  p_{\text{IBSO}}^{\pm\equat} = 2 \mp a + 2 \sqrt{1\mp a} \,.
\end{align}

\subsection{Polar IBSO}
\label{sec:polar-ibso}

Setting $x=0$ we get the factorization,
\begin{align}
  \label{eq:polar-ibso}
  I(x=0) = (p^4 - 4 p^3 + 2 a^2 p^2 + a^4)^{2}
  \,.
\end{align}
As this is the square of a quartic, there are four solutions by radicals.
The physical solution for $p_{\text{IBSO}}^{\text{pol}}$ is
\begin{align}
  p_{\text{IBSO}}^{\text{pol}} = {}& 1 + \sqrt{X_2} + \sqrt{3 - a^2 + \frac{2 - a^2}{\sqrt{X_2}} - X_2} \\
  X_{2} ={}& 1 - \frac{a^2}{3} + \frac{2 a^4}{3 X_{1}^{1/3}} + \frac{X_{1}^{1/3}}{6} \\
  X_{1} ={}&a^4\bigl(27 + \sqrt{27 (27 - 16 a^2)}\bigr) - 8 a^6 \,,
\end{align}
where all radical expressions are real in the physical region.

\subsection{Extremal IBSO}
\label{sec:extremal-ibso}

In the extremal limit $a\to 1$, we get the simplification
\begin{align}
  I(a=1) = (p-1)^{2} \Bigl(&
  (p-6 ) p^5 + p^4 + z_{m}^{4} \nn\\
  &{}+ 2 p^2 (1 + p) (-1 + 3 p) z_{m}^{2}\nn \\
  &{}+ p (2 + 9 p) z_{m}^{4} \Bigr) \,,
\end{align}
where $z_{m}^{2} = 1 - x^{2}$.
Here we see the same type of phenomenon as already discussed in
Sec.~\ref{sec:circ}.  For most values of $x$, one smoothly-varying
root of the sextic plays the role of $p_{\text{IBSO}}^{\text{ext.}}$.
But this root linearly crosses the constant root at $p=1$ at some
critical value of $x_{\text{kink}}^{\text{ext. IBSO}}$.  We find this
value by inserting $p=1$ into the sextic and thus have to solve $0=4
- 8 x^2 + 3 x^4$ to find the inclination of the kink.  We find the
location of the kink is
\begin{align}
  \label{eq:x-circ-extr-kink-IBSO}
  x_{\text{kink}}^{\text{ext. IBSO}} = \sqrt{2/3}
  \,,
\end{align}
which was previously found by other methods in~\cite{Hod:2017uof}.
Thus when $a=1$ and $x \ge  x_{\text{kink}}^{\text{ext. IBSO}} $, the
IBSO is at $p=1$. This kink behavior can be seen in
Fig.~\ref{fig:circ-ibso-extremal}.

\section{Numerical method for calculating the separatrix via connection to homoclinic orbits}\label{apdx:homoclinic_method}

In this appendix we generalize the approach of Ref.~\cite{Levin:2008yp} to generic orbits.
In general the radial equation has four distinct roots.
On the separatrix two of these roots coalesce so that $r_2 = r_3$.
Thus we can write the radial equation in the form
	\begin{align}
		R(r) &= -\beta(r-r_4)(r - r_2)^2(r - r_1)		\label{eq:R_factored_noneq}	
  \,,
\end{align}
where recall that $\beta = 1- \E^2$.
Comparing the coefficients of $r^2$ and $r^3$ in the above with the same coefficients in Eq.~\eqref{eq:radial_potential} and solving simultaneously for $\{r_1, r_2\}$ we find
\begin{align}\label{eq:rmax_noneq}
  r_1 = -r_2 + \frac{1 + \sqrt{1-\beta  \left(a^2 \beta +\L^2+\Q+2 r_2 \left(\beta  r_2-1\right)\right)}}{\beta }
  \,.
\end{align}
	We could now substitute $\{\E,\L,\Q\}$ with their values for spherical orbits with radius $r_2$ and substitute the result into Eq.~\eqref{eq:p_e_definitions}. This would give us a parametric equation for $\psep(r_0,x)$ and $\esep(r_0,x)$. Our goal is to find $\psep(a,e,x)$. To do this we need to numerically root find to get the solutions of $e = \esep(r_0, x)$. To do this stably across the entire parameter space we need to bracket the root. The value of $\esep(r_0)$ varies from $e=0$ when $r_0 = r_\isso$ to $e=1$ at $r_0 = r_\ibso$. Unfortunately, at $r_0 = r_\ibso$ the maximum orbital radius, $r_1$, diverges. This can be overcome by writing
	\begin{align}
		r_1 = \frac{r_1^\reg}{\gamma}
  \,,
	\end{align}
	where $r_1^\reg$ remains finite as $e\rightarrow 1$ and $\gamma\rightarrow 0$ as $e\rightarrow 1$. In formulating the equation for $r_1^\reg$ we have to be careful to avoid any divisions by $\beta = (1-\E^2)$ as $\E\rightarrow 1$ as $e\rightarrow1$. This is easily arranged and we find
\begin{align}
  &r_1^\reg = 2(\L-a\E)^2 + 2\Q \\
  &+ r_2^2 (-1 + r_2 \beta + \sqrt{1 - \beta(\L^2 + \Q + a^2\beta - 2r_2(1 - r_2\beta))}) \nonumber	\\
  & \gamma/r_2 +\beta  r_2 = \nonumber \\
  & 2 -2 \sqrt{1 + \beta(-a^2 \beta - \L^2 - \Q - 2 \beta r_2{}^2 + 2  r_2)}
  \,.
\end{align}
With these definitions we can define the eccentricity along the separatrix as
\begin{align}\label{eq:es_reg}
  e^\reg_\text{sep} &= \frac{r_1^\reg - r_2\gamma}{r_1^\reg + r_2\gamma}\\
  p^\reg_\text{sep} &= \frac{2 r_1^\reg r_2}{r_1^\reg + r_2 \gamma}
  \,.
  \label{eq:ps_reg}
\end{align}
In this equation you can directly substitute $e^\reg_\text{sep}$ and get $e^\reg_\text{sep}(r_\ibso) = 1$.

Putting it all together the algorithm for robustly locating the generic Kerr separatrix is
\begin{enumerate}
\item Pick a value for each of $\{a,e,x\}$.
\item Numerically solve $e^\reg_\text{sep}(r_2, x) = e$ by root finding between $r_2 = r_\ibso$ and $r_2 = 10$ (this is beyond any value the $r_\isso$ can take). In evaluating Eq.~\eqref{eq:es_reg} use the formula for $\{\E,\L,\Q\}$ for spherical orbits with radius $r_2$.
\item Compute $p^\reg_\text{sep}$ using Eq.~\eqref{eq:ps_reg}.
\end{enumerate}
The above algorithm was implemented into the \texttt{KerrGeodesics} \textsc{Mathematica} package in the Black Hole Perturbation Toolkit. Note this algorithm requires the location of the IBSO to be calculated beforehand. The Toolkit implementation found this by first locating the photon sphere radius, $r_\text{ph}$, and then root finding on $\E(r_\ibso) = 1$ noting that $r_\ibso > r_\text{ph}$ to bracket the root. This added two root finds to the process which slowed the algorithm down with respect to the new one presented in the main body of this article. The code in the Black Hole Perturbation Toolkit has now been upgraded to use the more efficient method.

Shortly after the above algorithm was implemented in the Black Hole Perturbation Toolkit Ref.~\cite{Rana:2019bsn} published their generalization of the approach in Ref.~\cite{Levin:2008yp}. 

\bibliographystyle{apsrev4-2}
\bibliography{references}

\end{document}